\documentclass[aps,reprint]{revtex4-2}
\usepackage[LY1]{fontenc}
\usepackage[utf8]{inputenc}
\usepackage{mathrsfs}
\usepackage{mathtools}
\usepackage{physics}
\usepackage{tabularx}
\usepackage{amsmath}
\usepackage{amsfonts}
\usepackage{bm}
\usepackage{framed}
\usepackage{graphicx}
\DeclareUnicodeCharacter{2212}{-}
\usepackage{enumitem}
\usepackage{blindtext}
\usepackage[colorlinks, breaklinks, urlcolor={blue}, linkcolor={red}, citecolor={blue}]{hyperref}
\usepackage{qcircuit}

\usepackage[colorinlistoftodos, color=green!40, prependcaption]{todonotes}
\usepackage{amsthm}

\usepackage{xcolor}

\usepackage{adjustbox}
\usepackage{placeins}
\usepackage[T1]{fontenc}
\usepackage{lipsum}
\usepackage{csquotes}

\draft

\begin{document}

\title{Scaling Network Topologies for Multi-User Entanglement Distribution}
\author{Muhammad Daud}
\email{m.daud.sns@gmail.com}% Your name
\affiliation{School~of~Natural~Sciences,~National~University~of~Sciences~and~Technology,~Islamabad,~Pakistan.}

\author{Aeysha Khalique}
\email{aeysha.khalique@sns.nust.edu.pk}
\affiliation{School~of~Natural~Sciences,~National~University~of~Sciences~and~Technology,~Islamabad,~Pakistan.}
\affiliation{National~Centre~for~Physics~(NCP),~Shahdra~Valley~Road,~Islamabad~44000,~Pakistan.}

\date{\today}

\begin{abstract}
Future quantum internet relies on large-scale entanglement distribution. Quantum decoherence is a significant obstacle in large-scale networks, which otherwise perform better with multiple paths between the source and destination. We propose a new topology, a connected tree, with a significant number of redundant edges to support multi-path routing of entangled pairs. We qualitatively analyze the scalability of quantum networks to maximum user capacity in decoherence for different topologies. Our analysis shows that thin-connected tree networks can accommodate a larger number of user pairs while maintaining a high-routing environment, resulting in less dependence on quantum memories for routing than distributed lattice or P-2-P topologies. Thus, leading to robustness against decoherence and better key generation rates among multiple communicating parties in quantum key distribution. 
\end{abstract}

\maketitle
\section{Introduction}\label{1}

Technological advancements based on quantum theory are advancing in the second quantum revolution. With progress in quantum key distribution~\cite{BENNETT20147}, quantum computation~\cite{nielsen_chuang_2010}, quantum transducers~\cite{Lauk_2020} and multiplexers~\cite{8750173} along with quantum memories, it is not a far fetched dream that we will be looking at a large scale quantum internet~\cite{Kimble2008}. The preliminary task for such a quantum internet will be to build a network that can distribute quantum resources to anyone seeking to implement any desirable protocol. Quantum networks exploit many techniques to achieve aforementioned tasks, like trusted relays and optical switching~\cite{Chen2021, Stucki_2011, Peev_2009}, quantum repeaters~\cite{Ruihong_2019} or quantum teleportation-based network~\cite{PRXQuantum.1.020317}. Entanglement has proven to be a helpful quantum resource that can achieve unprecedented results for global quantum communication and computation~\cite{rohde_2021}.

Entanglement distribution networks can become the backbone for scalable quantum networks~\cite{doi:10.1126/science.1140300, Yuan2008} because it uses entanglement swapping and quantum memories to demonstrate the sharing of entangled pairs to distant parties. It is similar to classical data delivery networks that rely on transmission control protocol~(TCP), which focuses on the delivery of classical bit packets that can be used to share information~\cite{rohde_2021}. For scalable classical networks, it is an important consideration that each pair of source and destination can send/receive data packets, therefore, TCP employs a \textsc{REPEAT-UNTIL-SUCCESS} strategy. However, if a larger number of user pairs wish to share information at a given time, it will result in congestion, leading to low latency and bad bandwidth. Classically, a solution to this problem is proposed as multi-path data routing, where each user pair wishes to use more than one path to route their data packets~\cite{KHEIRKHAH2019106852}. The \textsc{REPEAT-UNTIL-SUCCESS} strategy doesn't hinder entanglement distribution networks. These networks share entangled bits with no prior information between user pairs to be used for various protocols.

Entangled pairs from different paths incur decoherence, which scales exponentially as the network grows, posing serious limitations on large-scale quantum networks. Quantum repeaters based on entanglement swapping which utilizes quantum memories and entanglement purification to produce high-quality entangled pairs~\cite{7010905} can be used to share entanglement at will~\cite{Sun2016, Autebert_2016, Herbauts:13, 9748401, Inagaki:13, doi:10.1126/science.aan3211}. However, a significant challenge towards a scalable quantum internet is effectively providing entangled pairs for multiple user pairs. Though it is obvious, the quantum internet much like the classical internet will lack structure at large, and quantum memories will be an integral part of it. Both ideas are far from realization due to technological and strategic reasons.
Nevertheless, it is important to find short-term solutions to minimize the use of low-performing quantum memories and promote large-scale quantum networks. One way to do this is to introduce redundancy in the network to allow multi-path routing of entangled bits. One of the main obstacles in achieving this is the inherent quantum complexities like no-cloning theorem, decoherence, and error estimation due to the use of logical qubits~\cite{leone2021qunet}. The second challenge is optimal path-finding because of our goal to minimize the decoherence cost for each source and destination~\cite{PhysRevLett.120.080501} so that the whole network can benefit from it.

We propose using network topologies with higher connectivity and link redundancy to ensure minimum competition and optimized path-finding for quantum networks. It has been previously shown that sharing entanglement among all network parties is beneficial~\cite{PhysRevLett.120.080501}. Our proposed topology for a city-wide node of future quantum internet incentivizes every user pair to avoid congestion and maximize multi-path routing of entangled pairs. In this work, we address the issue of the scalability of quantum networks by adding redundant paths and analyzing the scaling of efficiency-fidelity trade-offs in different topologies. The study takes many scenarios into account like routing with or without quantum memories and routing with or without redundant edges or both.

We present a qualitative analysis of network topologies simulated at arbitrarily fixed parameters. Our analysis shows that topological consideration significantly improves network statistics for multiple user pairs. For a tree topology with rings of redundant edges at each level~(connected tree), we show that the edge scheme, defined via branching parameter and depth, greatly enhances user accommodation and path purification thresholds. We also show through a comparison of a connected tree and lattice topology that a connected tree is more favorable for a scalable quantum internet in terms of accommodating more user pairs and providing more paths to them. The significance of topological considerations is shown in this work by simulating QKD protocol over both lattice and connected tree topologies at various user competitions.

In this article, firstly, Sec.~\ref{2} gives a detailed background study on cost vector analysis, the inherent differences of network topologies, and path-finding strategies. Secondly and most importantly, our findings are presented in Sec.~\ref{3} about the impacts of edge schemes and network topology on network statistics and QKD secret key rates and memory-less practical QKD network. Finally, we conclude in Sec.~\ref{4}.

\section{Background}\label{2}
To effectively distribute entanglement within quantum repeater networks, a series of essential tasks must be accomplished~\cite{rohde_2021}. Firstly, it is necessary to establish entanglement between adjacent parties, ensuring the network is interconnected and capable of transmitting quantum information. Secondly, entanglement swapping is utilized to entangle non-adjacent parties, expanding the network's reach and potential capabilities. Finally, entanglement purification techniques are employed to improve the quality of the entanglement, thereby increasing the accuracy and reliability of the transmitted information.

In this section, we provide the necessary background on various aspects of quantum networks and their analysis, viz. cost vector analysis in Sec.~\ref{CVA}, and network topologies in Sec.~\ref{compNet}. We introduce the routing strategies used in our work in Sec.~\ref{2D}

\subsection{Cost vector analysis}\label{CVA}
In quantum internet, quantum states dissipate while traversing through different channels, leading to a cost associated with each channel. Cost vector analysis (CVA) is based on dealing with the cost incurred while traversing quantum channels, without keeping track of the actual physical process. In cost vector analysis, all costs should be positive, additive, and bounded by triangular inequality~\cite{rohde_2021}. Quantum channels are intrinsically different from classical channels. They are not necessarily commutative, and their inherent probabilistic nature may not allow them to accumulate additively. 

The two most important costs associated with the quantum channels are loss and dephasing, which are commutative, yet multiplicative. The $i^\text{th}$ channel loss with transmissivity $\eta_i$ acting on the incident state $\hat\rho$ maps the state to~\cite{nielsen_chuang_2010},
\begin{align}\label{losschannel}
    \mathscr{E}^i_\text{loss}(\hat{\rho})=\eta_i\hat{\rho}+(1-\eta_i)\ket{\text{vac}}\bra{\text{vac}},
\end{align}
where the state $\ket{\text{vac}}\bra{\text{vac}}$ is the vacuum state with zero photons in it. 
The net transmissivity for $n$ channels, each of length $L_i$, is multiplicative
\begin{align}\label{nettrans}
\nonumber\eta_{\text{net}}&=e^{-\alpha L_{\text{net}}}\\&\implies e^{-\alpha (L_1+L_2+L_3+\cdots+L_n)}\implies\eta_1\eta_2\eta_3\cdots\eta_n,
\end{align}
where $\alpha$ is the characteristic of the optical fiber material. 

Dephasing causes the state to randomize its phase. The action of each $i^{th}$ dephasing channel with dephasing probability $p_i$ is~\cite{leone2021qunet},
\begin{align}\label{Dephaschannel}
    \nonumber\mathscr{E}^i_\text{dephasing}(\hat{\rho})&=p_i\hat{\rho}+(1-p_i)\frac{\hat{\rho}+\hat{Z}\hat{\rho}\hat{Z}}{2}\nonumber\\&=\big(\frac{1+p_i}{2}\big)\hat{\rho}+\big(\frac{1-p_i}{2}\big)\hat{Z}\hat{\rho}\hat{Z},
\end{align}
The dephasing channel is often parameterized as a pure phase-flip channel. When it is applied to one-half of the maximally entangled state, leaves the error state as $\hat{Z}\hat{\rho}\hat{Z}$, which is orthogonal to input state $\hat{\rho}$. We can therefore identify $p_i$ as the overlap probability of input and output state.
\begin{align}\label{Pnet}
P_{\text{net}}=\Pi_{i=1}^n\big(\frac{1+p_i}{2}\big).
\end{align}

We can make the cost of transmissivity $c_{\text {loss}}$ and dephasing $c_Z$ additive, to align these with the definition of cost vector, by taking the logarithm of $\eta_{\text{net}}$ and $P_{\text{net}}$
\begin{align}    c_\text{loss}=-\sum_{i=1}^n\log(\eta_i)&&c_{Z}=\sum_{i=1}^n\log\big(\frac{1+p_i}{2}\big)
\end{align}
Using such a technique not only allows us to deal with a quantum channel in a classical manner but also helps to keep track of the effectiveness of paths in large networks.

The states transmitted are the entangled pairs distributed through entanglement swapping~\cite{PhysRevLett.80.3891} from one channel to the next. Entanglement swapping being heralded is done until successful, so it does not affect the cost vector. However, entanglement purification~\cite{purificationPRL.76.722} is done after each swapping to increase the fidelity of weakly entangled pairs. Thus fidelity is updated to $F'_i$ after traversing through each channel,
\begin{align}
    F^{'}_i=&\frac{F_1F_2}{F_1F_2+(1-F_1)(1-F_2)}.
\end{align}
where $F_i$ is the fidelity of input entangled pairs given by $F_i = ( 1+p_i )/2$. $\eta_i$ being the efficiency of traversing $i^{th}$ channel, needs to be updated accordingly, taking into account whether the swapping and purification processes are feasible or not. The updated efficiency of $i$th channel is~\cite{leone2021qunet},
\begin{align}\label{6}
    \eta'_i=&\eta_1\eta_2[F_1F_2+(1-F_1)(1-F_2)]
 \end{align}
where $\eta_i$ is the loss probability for input entangled pairs. Thus, a channel is traverseable if and only if $F_i>1/2$. In the cost vector, there are only two parameters to keep track of, efficiency and fidelity.

\subsection{Scaling of Quantum Networks Topologies}\label{compNet}
Networks, in mathematical terms, are graphs of $V$ vertices and $E$ edges~\cite{rohde_2021}
\begin{align}
    G=(V,E).
\end{align}
The topology of a network is defined as the arrangement of vertices $V$ and how $E$ edges connect them. In quantum networks, each graph vertex behaves as a quantum node. It can be either an end user or a repeater node performing entanglement swapping and purification. An edge is a channel between two adjacent quantum nodes. These channels could be optical fiber links, free space, or satellite links. The transmission of a quantum resource throughout a network heavily relies on its topology. There are various network shapes with their advantages and disadvantages. We introduce the three used in our analysis.
\begin{figure}[ht!]
    \centering
    \begin{tabular}{c c c}
         (a)\includegraphics[width=38mm,height=40mm]{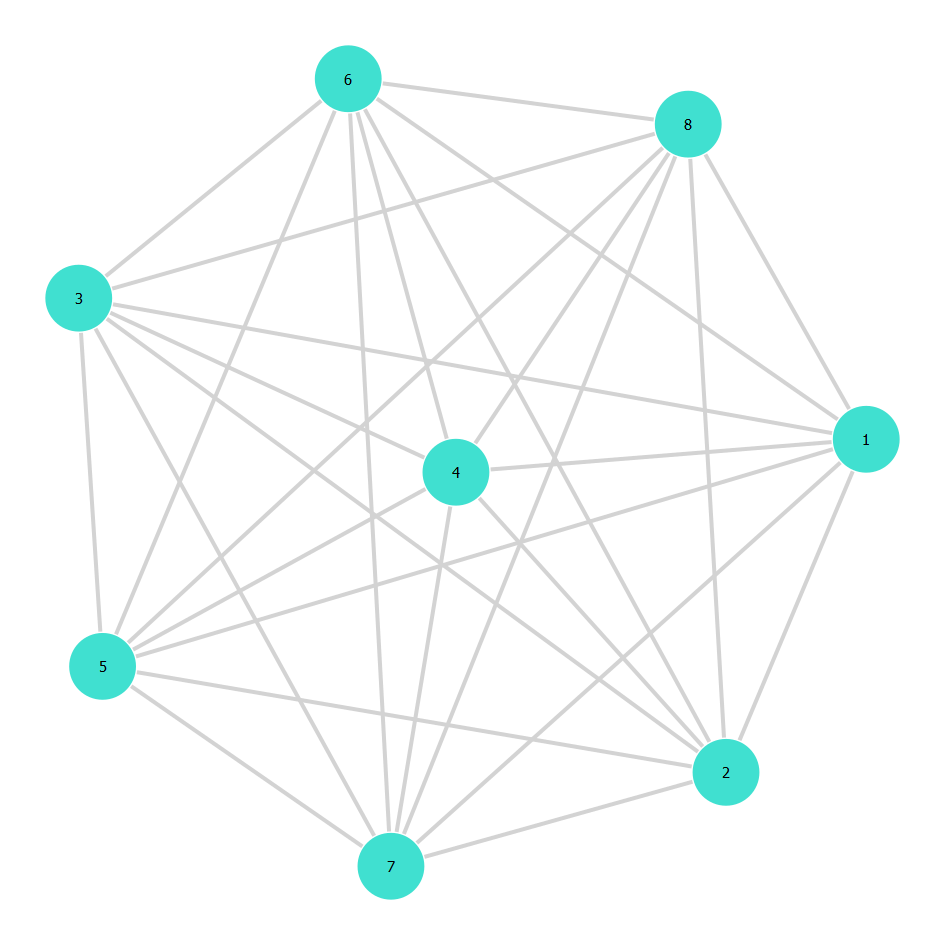}&(b)\includegraphics[width=38mm,height=40mm]{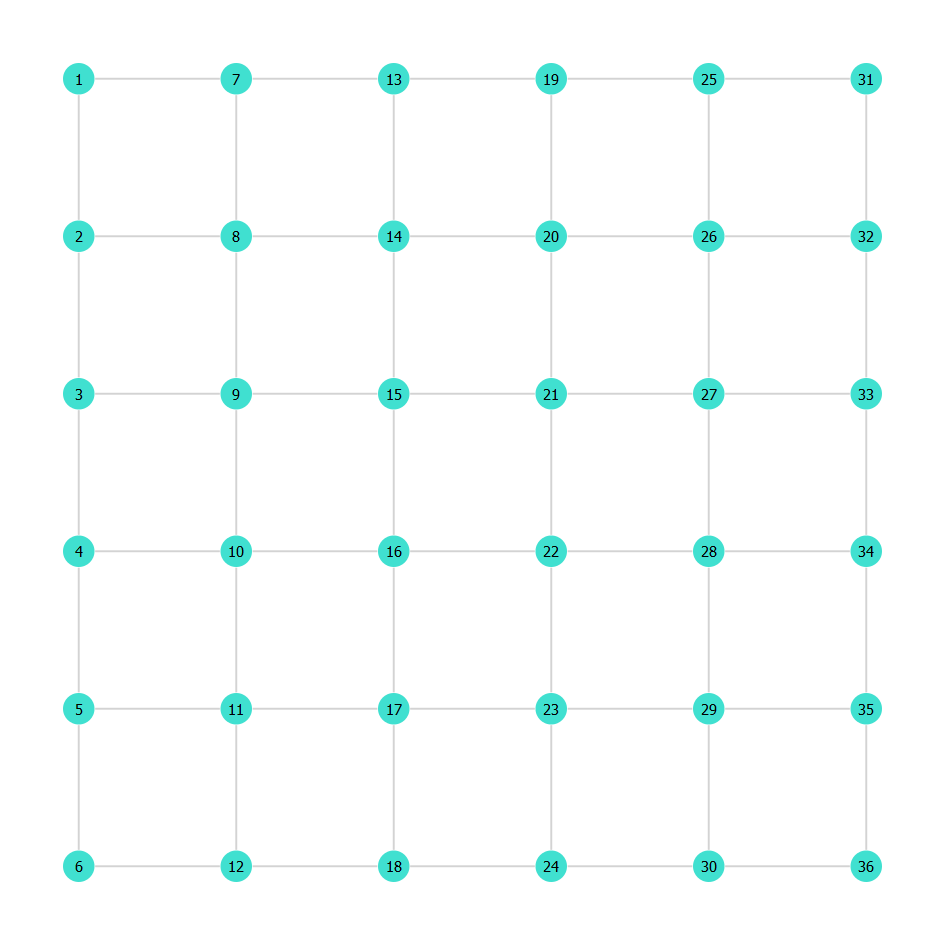}
         \end{tabular}
         (c)\includegraphics[width=40mm,height=40mm]{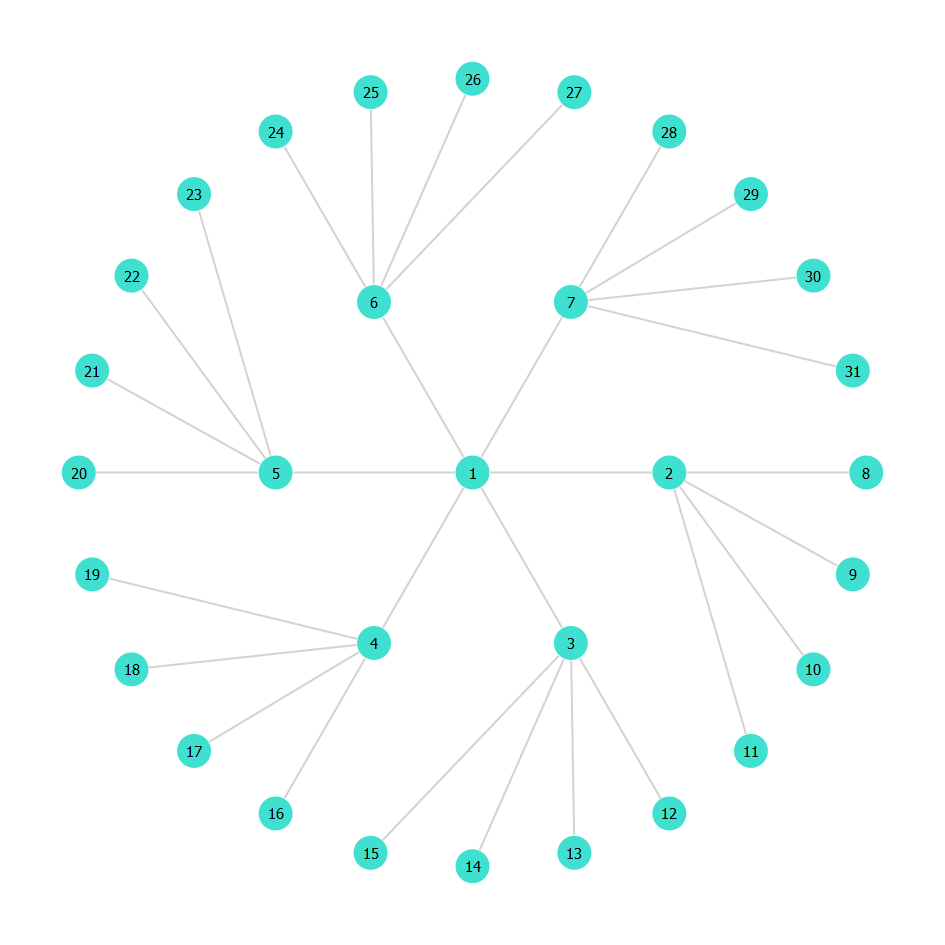}
    
    \caption{Examples of some network topologies, (a) A complete graph of 8 vertices, (b) A $6\cross 6$ lattice graph, (c) A balanced tree graph with branching parameters of 6 and 4 at depth 1 and 2 respectively.}
    \label{CGLATT}
\end{figure}

A complete graph shown in Fig.~\ref{CGLATT}(a) offers maximum connectivity as every node is linked to each other~\cite{rohde_2021}. This lessens the need for quantum memories and makes path-finding effortless. However, this topology is impractical for large-scale networks as the number of edges $|E|$ scales quadratically with the number of nodes $|V|$,
\begin{align}
    |E|=\frac{|V|^2 -|V|}{2} && \text{diameter}=d=1,
\end{align}
resulting in high infrastructure costs. 

 Fig.~\ref{CGLATT}(b) illustrates the lattice topology, a $m\cross n$ dimension grid, whose edges increase linearly with the number of vertices in each dimension. The lattice topology is advantageous for accommodating multiple user pairs because it offers numerous routing options. However, it has a significant drawback in network optimization. As the network size increases, the shortest path becomes longer, which increases the path cost significantly.~\cite{rohde_2021},
\begin{align}
    |E|=m(n-1)+n(m-1)&& d=\mathcal{O}(m+n).
\end{align}

A tree topology shown in Fig.~\ref{CGLATT}(c) is an elementary topology characterized by depth and branching parameters. They are inherently acyclic and have only one path between two nodes. Its edges increase linearly with the number of vertices, but diameter scales logarithmically~\cite{rohde_2021},
\begin{align}\label{treedist}
    |E|=|V|-1 && d=\mathcal{O}(\log(|V|)).
\end{align}
A major disadvantage of tree topology is its vulnerability to failure, which makes it unsuitable for multi-user quantum networks. On the other hand, thin trees have proven beneficial for studying path optimization algorithms like asymmetric traveling salesman problems~\cite{Haji2018-lz}.

\subsection{Path-finding strategies}\label{2D} 
The search for an efficient quantum network relies on finding the shortest path with the least cost. The cost vector analysis reduces the search to utilize classical shortest-path algorithms. We utilize \href{https://github.com/drpeterrohde/QuNet/tree/main}{QuNet}, a quantum network simulator, for our simulations. It relies on the Dijkstra shortest path algorithm, introduced in 1959~\cite{Dijkstra1959} to determine the shortest path in any given graph. The algorithm involves making multiple decisions to identify the path with the lowest cost. For example, in Fig.~\ref{1,2 path routing}, if we begin at source point `A', the algorithm will search for the path with the least cost and follow it to the next node, repeating the process until it reaches destination `B'. The algorithm has a maximum run-time of $\mathcal{O}(|V|^2)$. It has been improved with a run time of $\mathcal{O}(|E|+|V|\log(|V|))$~\cite{rohde_2021}.
Dijkstra's shortest path algorithm takes a heuristic approach towards path-finding between single-user pairs. Therefore when we scale this algorithm to multiple users, the problem becomes a vehicle routing problem which is a known NP-hard problem~\cite{rohde_2021}. Hence, a complete multi-user optimization is optional.
\begin{figure}[!ht]
    \centering
    \includegraphics[height=40mm,width=40mm]{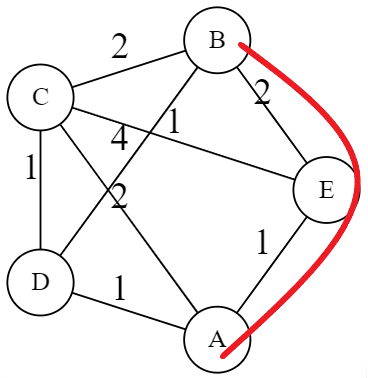} 
    \caption{Shortest path routing manifesting in a network.}
    \label{1,2 path routing}
\end{figure}

In traditional internet using a single path utilizes User Datagram Protocol (UDP) and the repeat-until-success method to ensure packet distribution~\cite{rohde_2021}. However, multiple paths in classical routing are of particular interest because they can make the network more resilient towards attacks by improving end-to-end data delivery~\cite{CMP}. In quantum internet, the guarantee of packet distribution is associated with the fidelity of the entangled pairs shared between end users. Multi-path routing aims to use more than just the shortest path to maximize the network's ability to support entanglement purification. This approach generates high-fidelity Bell pairs by purifying Bell pairs from different routes~\cite{doi:10.1116/1.5119961, PhysRevLett.120.080501,leone2021qunet}. All the users of the network know which Bell state to
prepare and each node can make any number of those Bell states deterministically. Since we are only comparing the routing costs, which are incurred after a healthy
entangled pair has traversed through a channel, the losses represent the true routing cost. With the repeat-until-success strategy~\cite{KHEIRKHAH2019106852}, a quantum internet ensures a flawless transmission of entangled pairs. 

\begin{figure}[ht!]
    \centering
    \begin{tabular}{c}
         (a)\includegraphics[height=35mm,width=35mm]{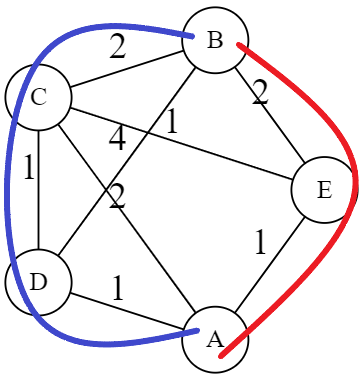}\\(b)\includegraphics[height=45mm,width=60mm]{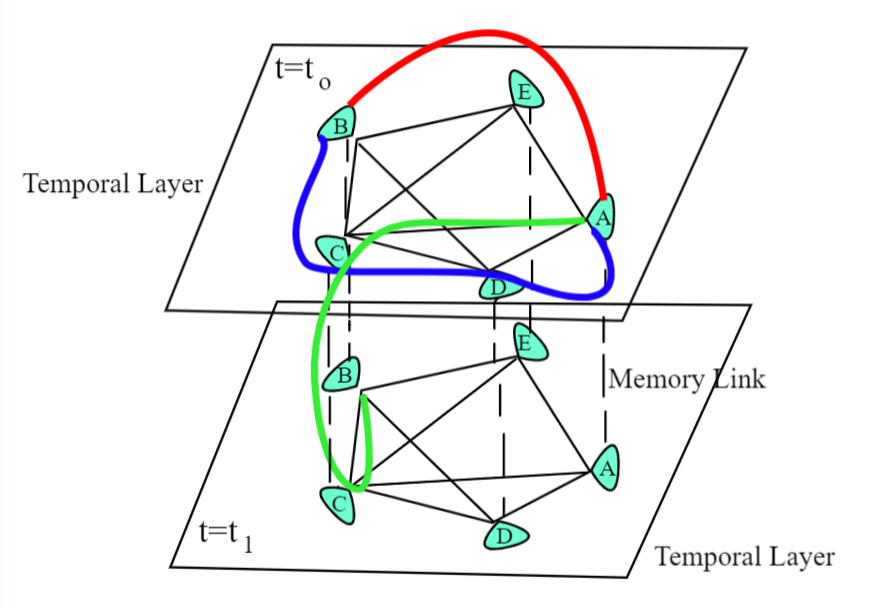}
    \end{tabular}
    \caption{(a) Multi-path routing strategy, (b)Temporal multi-path routing strategy manifesting in a network.}
    \label{3 path routing}
\end{figure}

The multi-path routing algorithm takes a greedy approach, by considering all available paths simultaneously and using Dijkstra's algorithm to find the shortest path that can be utilized in each iteration. The algorithm removes the utilized path from the network to avoid repetition in subsequent iterations. Fig.~\ref{3 path routing}(a), shows two different paths between nodes `A' and `B', with the red being the shortest and the blue being the next shortest.
The multi-user algorithm is a more generalized form of the multi-path algorithm~\cite{leone2021qunet}, aiming to maximize the graph's utilization. The network graph $G$ is partitioned into $n$ sub-graphs, $G_i$ for $i\in\{1,\dots,n\}$. Each sub-graph contains a source node, a destination node, and all the paths used between them while adhering to specific rules,
\begin{align}
    G=\bigcup_{i=1}^{n}G_i \quad,\quad G_i\cap G_j=\emptyset \quad,\quad \forall\{i\neq j\}.
\end{align}
Quantum memory being an integral part of Entanglement distribution is simulated as an edge in time. Thus, temporal routing guarantees a path between each user pair by routing entanglement at different times when the path is available. An asynchronous routing scheme~\cite{leone2021qunet} is given in Fig.~\ref{3 path routing}(b). Here, `A' and `B' establish entanglement using three paths. The entangled pairs are transmitted at some instance $t_0$ through red and blue paths, but no more paths are available. Temporal routing allows the entangled pairs to traverse to another temporal layer at time $t_1$. Here the link between `C' and `B', which is clogged up in the blue path, is freed to be used in the green path. Thus, guaranteeing a path at the expense of higher costs.

This approach does not require tracking of qubits and hence bypasses some quantum complexities. It only needs bookkeeping of two parameters as costvector because with every channel that is traversed the fidelity and efficiency of entangled pairs changes, as discussed in Sec.~\ref{CVA}, establishing a link that takes swapping and purification into account. The optimal route between end users is based on the minimization of the cost vector.

\section{Scaling for different network topologies for multi-path in multi-user networks}\label{3}
The present section presents our research findings on quantum network analysis, an area of paramount importance in quantum information. Available quantum relay networks are primarily based on two topologies, namely Point-to-Point~(P2P) and Star. The problem with such topologies is that they do not support multi-path routing entanglement. The closest contender widely studied is lattice topology. It is not an effective choice because when multiple user pairs join the quantum network, the path-finding becomes competitive, and paths in lattice scale bi-linearly in $X$ and $Y$ dimensions. We propose a topology that is not only suitable for large-scale quantum repeater networks but is relatively effective when scaled to high-user competition scenarios. We call it a "connected tree", which incorporates redundant rings of edges at each level of the existing tree network to support multiple user pairs. The connected tree has twice as many edges as the associated tree graph, with $|E|=2(|V|-1)$. By overcoming the tree network's acyclic nature limitation, the connected tree provides greater flexibility for accommodating user pairs and enables more paths between users. Specifically, our study highlights the fundamental role that network topologies play when scaling quantum networks for protocols, such as multi-user quantum key distribution and distributed quantum computing.

Our research showcases the proof of concept for scalability and multi-path routing in tree topologies, as discussed in Sec.~\ref{3A}. Furthermore, we demonstrate in Sec.~\ref{3B} the significant impact that edge schemes of topology have on network statistics. In Sec.~\ref{3c}, we compare the effect of network topologies on quantum network statistics. In Sec.~\ref{3D}, we exhibit the influence of the multi-user routing scheme on quantum key distribution networks with various topologies. Lastly, in Sec.~\ref{3E}, we show an advantage of redundant edges for a practical memory-less QKD network to P-2-P QKD network.

\subsection{Proof of principle of adding redundant edges in tree networks}\label{3A}
Proof of principle emphasises the importance of redundant edges in tree topology and the need for quantum memories to accommodate multiple user pairs in the network. The acyclic nature of the tree graph makes it a good candidate for single-path routing between a user pair, but all the nodes in that single will now only act as repeater nodes. This leads to the catastrophic failure of tree topology in multi-user competition. Adding redundant edges to the topology in a specific manner, i.e. a ring of edges connects each node at a given depth, will allow a user pair to opt for multi-path routing of entangled pairs. These redundant edges remove the tree topology's acyclic nature and help accommodate more user pairs to support multi-user routing. Previously, it has been shown for lattice networks~\cite{leone2021qunet}, and here it is extended to tree networks.

Tree networks only have one path between any two nodes, causing simple routing algorithms like Breadth First Search (BFS) and Depth First Search (DFS) to find the shortest path efficiently.
\begin{figure}[!ht]
    \centering
    \begin{tabular}{c c}
         (a)\includegraphics[height=36mm,width=38mm]{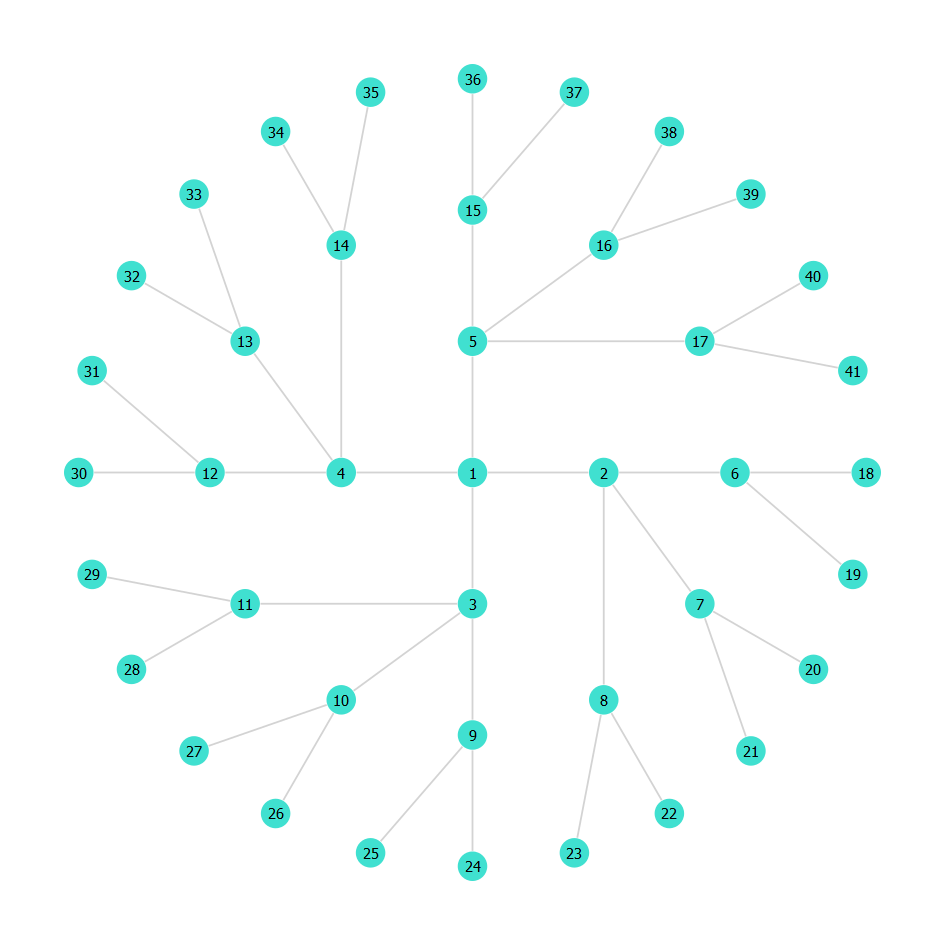} & (b)\includegraphics[height=36mm,width=38mm]{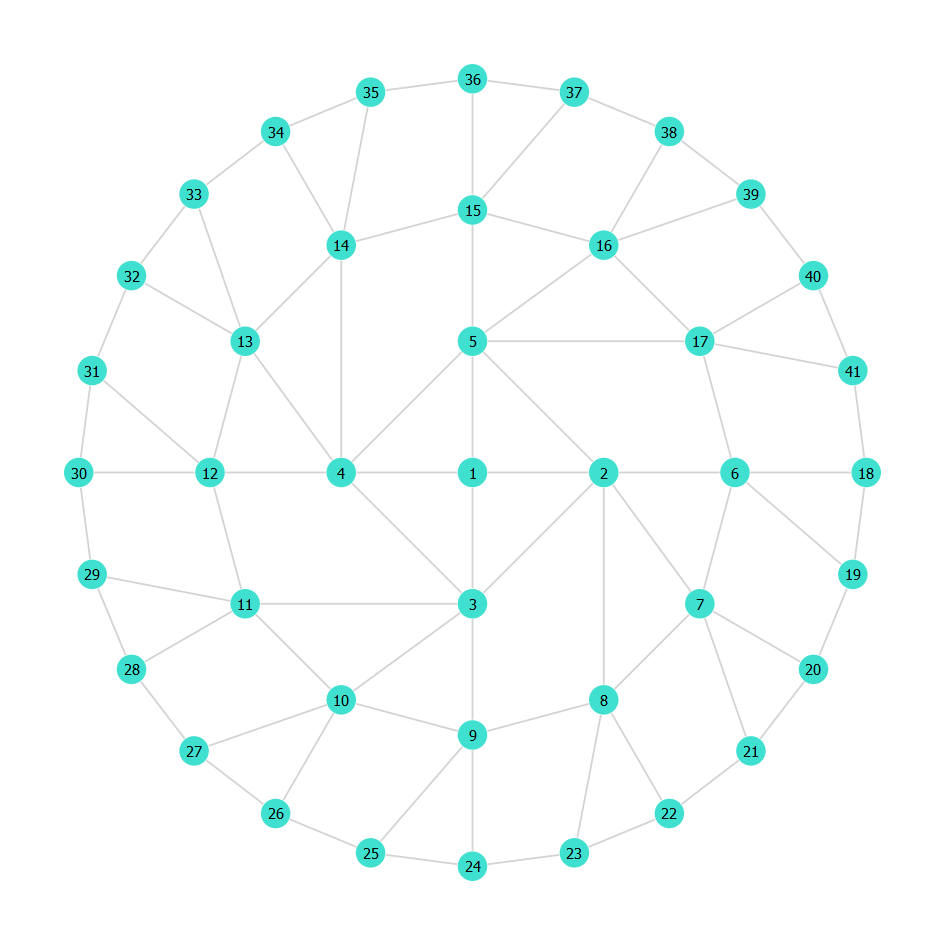}
    \end{tabular}
    \caption{A graphical representation of tree topologies. (a) a balanced tree with three depth levels with a branching ratio decreasing at each depth, (b) a connected tree made by adding concentric rings of edges at each depth.}
    \label{trees}
\end{figure}
Path-finding in tree graphs is relatively easy for a single-user pair. However, as we scale the graph, the shortest path-finding problem becomes a vehicle rescheduling problem~\cite{rohde_2021}. A connected tree topology is used to overcome it, which can accommodate more user pairs by providing more routing options.
\begin{figure}[!ht]
    \centering
    \begin{tabular}{c c}
         (a)\includegraphics[height=34mm,width=38mm]{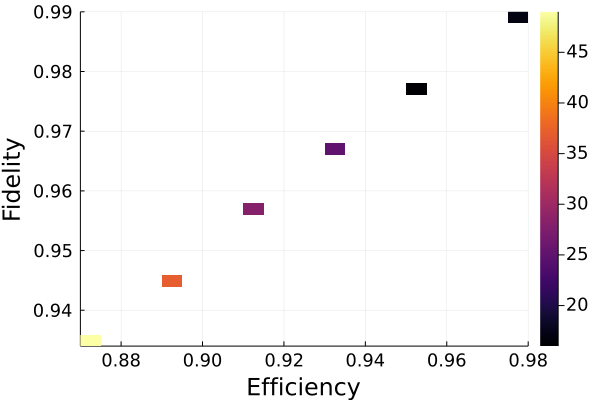} & 
         (b)\includegraphics[height=34mm,width=38mm]{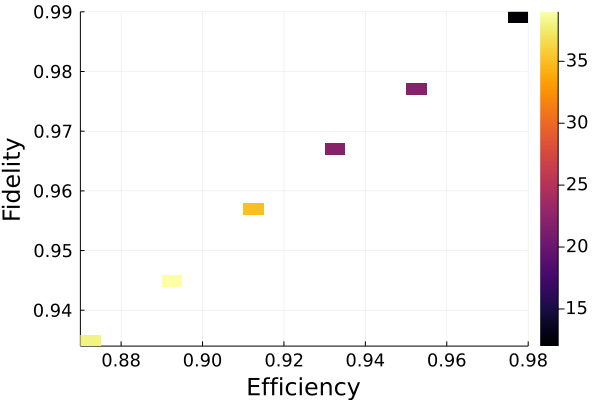}\\
         (c)\includegraphics[height=34mm,width=38mm]{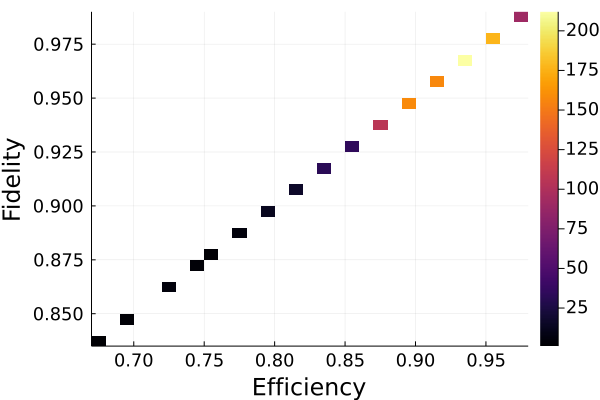} & 
         (d)\includegraphics[height=34mm,width=38mm]{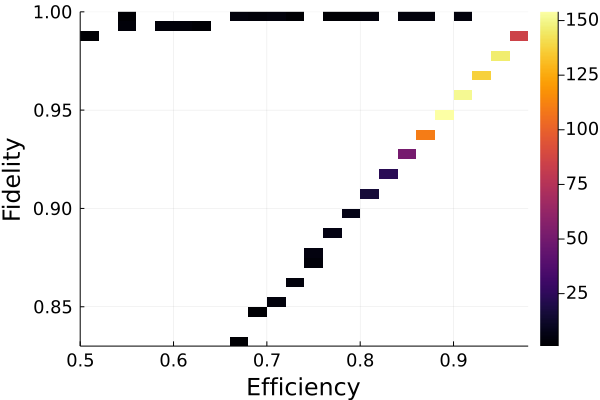}
    \end{tabular}
    \caption{Trade-off is shown between fidelity of entangled pairs and the efficiency of the path/paths used to share pairs between source and destination. (a), (c) and (b), (d) shows the single path and Multi-path routing respectively. {\color{red} The color scale represents the bin count for viable routes
utilized out of the 2000 routing opportunities} }
    \label{simple tree routing}
\end{figure}
41 node tree topologies are shown in Figs.~\ref{trees}(a) and (b). Both networks are simulated with similar edge costs of $0.1$~dB depicting loss and dephasing channels. QuNet is used to find paths between $20$ user pairs for $100$ trails. The only variable parameter for proof of principle is the number of allowed paths between each user pair.

Figure~\ref{simple tree routing} illustrates routing data for the shortest path and two paths of entangled pairs. QuNet ran the simulation 2000 times. The color scale represents the bin count for viable routes utilized out of the 2000 routing opportunities. A given box on the graph means that some routes have efficiency and fidelity associated with that box, and the color of the box tells us how many viable routes have those specific efficiencies and fidelities. As shown in Figs.~\ref{simple tree routing}(a) and (b), the feasible paths in tree topology are scarce. The linear distribution of path cost data confirms that the tree fails to provide multiple paths and accommodate multiple users. Whereas from Figs.~\ref{simple tree routing}(c) and (d), it can be extrapolated that connected tree provides a considerably larger number of routing options than trees. The connected tree topology can also accommodate more user pairs because of the removal of acyclic bound found in a trivial tree topology. In Fig.~\ref{simple tree routing}(d), apart from the linear distribution of path data, these outliers represent the use of two paths. The trade-off between the fidelity and efficiency is highlighted by these outliers, where we can see that there are a few routing options, that have low efficiencies meaning multi-path routing has incurred greater transmission loss, but the fidelities are high meaning purification has allowed the user pair to retrieve better entangled pairs. Thus the trade-off is whether to bear low-efficiencies for high fidelities. Thus topology is a key component of this trade-off. The advantage of multi-path routing arises from entanglement purification, where a gain in fidelity is achieved at the expense of decreased efficiency. The simple trade-off that is implied by this
comparison is that there is an optimal arrangement of edges that keeps track of user pairs so a minimum number of edges are traversed and at the same time allowing multi-path routing between maximum user pairs.

\subsection{Impact of Edge Scheme on Scalability}\label{3B}
For any global quantum server, scalability is a great challenge. We explore the role of the edge scheme of topology in network statistics, i.e. performance or path-finding in user competition.

\begin{figure}[!ht]
    \centering
    \begin{tabular}{c c}
         (a)\includegraphics[width=30mm,height=30mm]{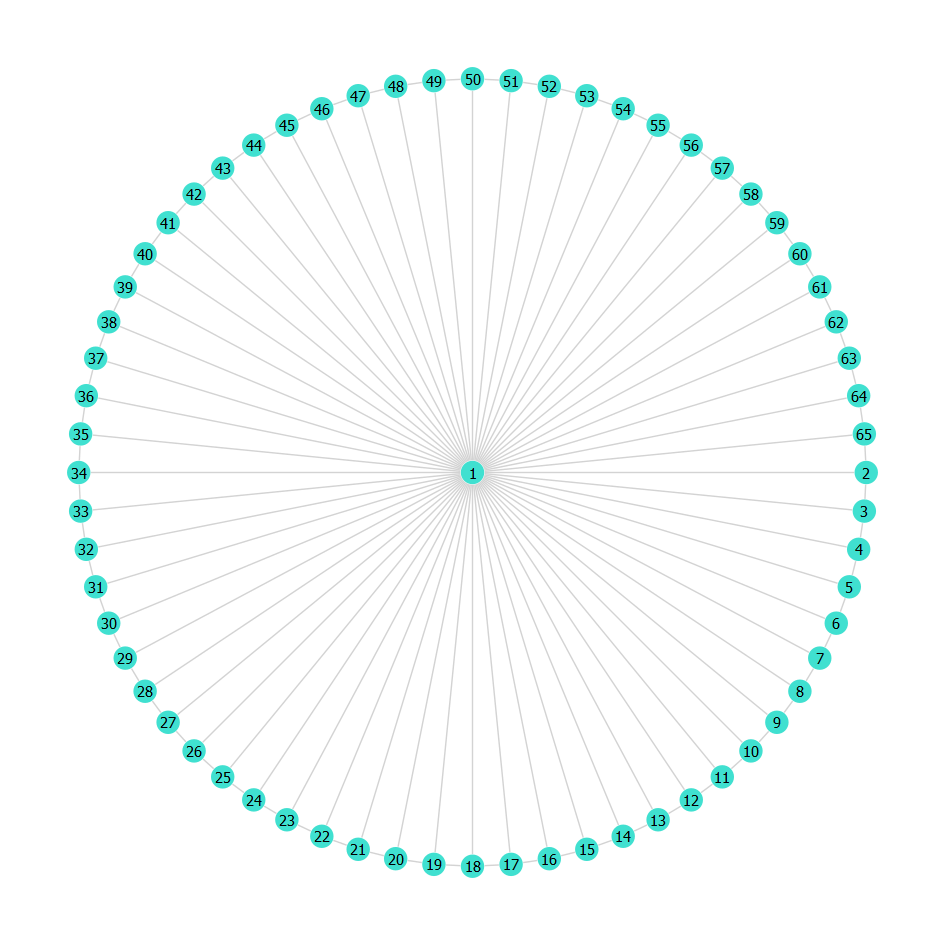}&(b) \includegraphics[width=40mm,height=30mm]{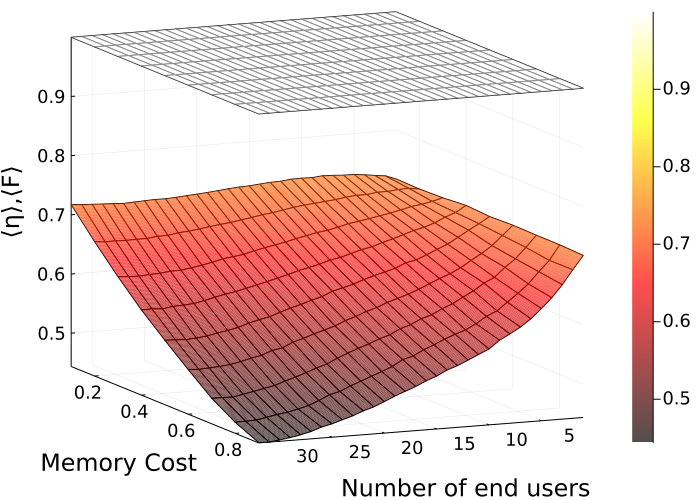}
    \end{tabular}
         (c)\includegraphics[width=40mm,height=30mm]{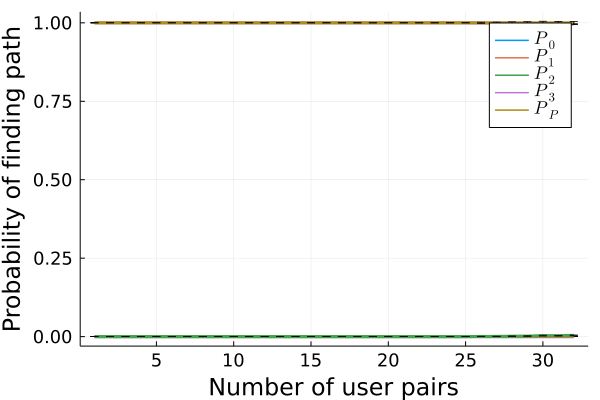}
    \begin{tabular}{c c}
         (d)\includegraphics[width=30mm,height=30mm]{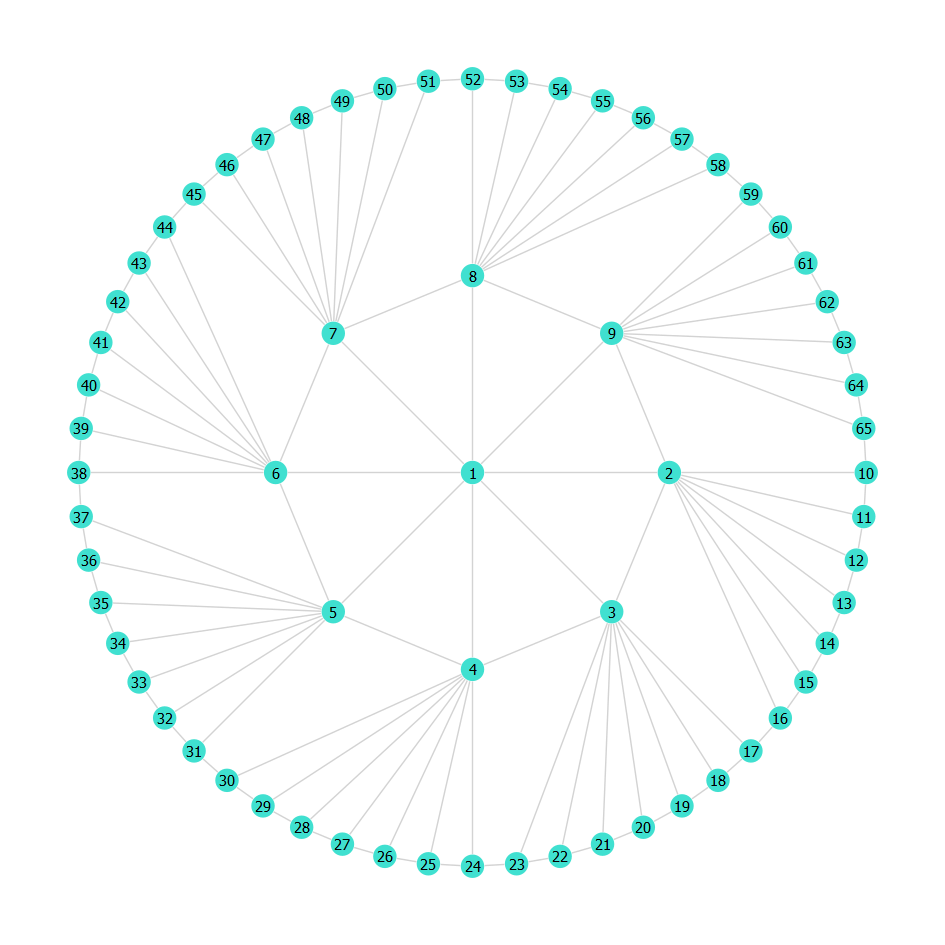}&(e) \includegraphics[width=40mm,height=30mm]{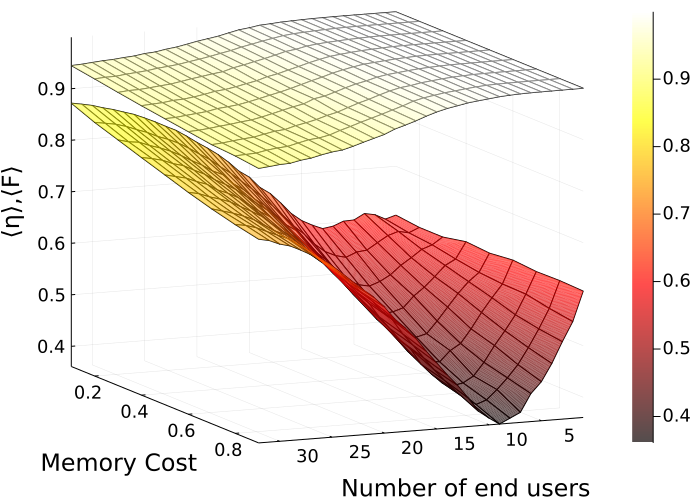}
    \end{tabular}
         (f)\includegraphics[width=40mm,height=30mm]{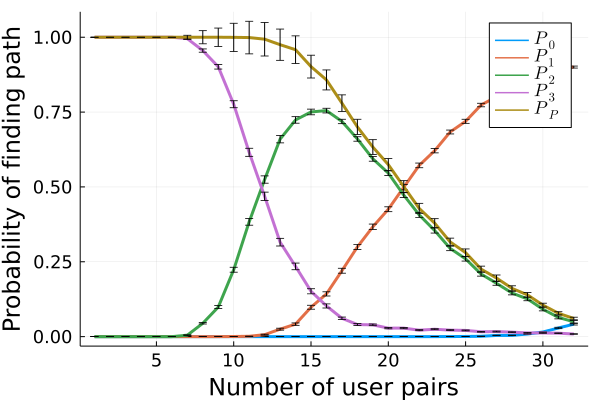}
    \begin{tabular}{c c}
         (g)\includegraphics[width=30mm,height=30mm]{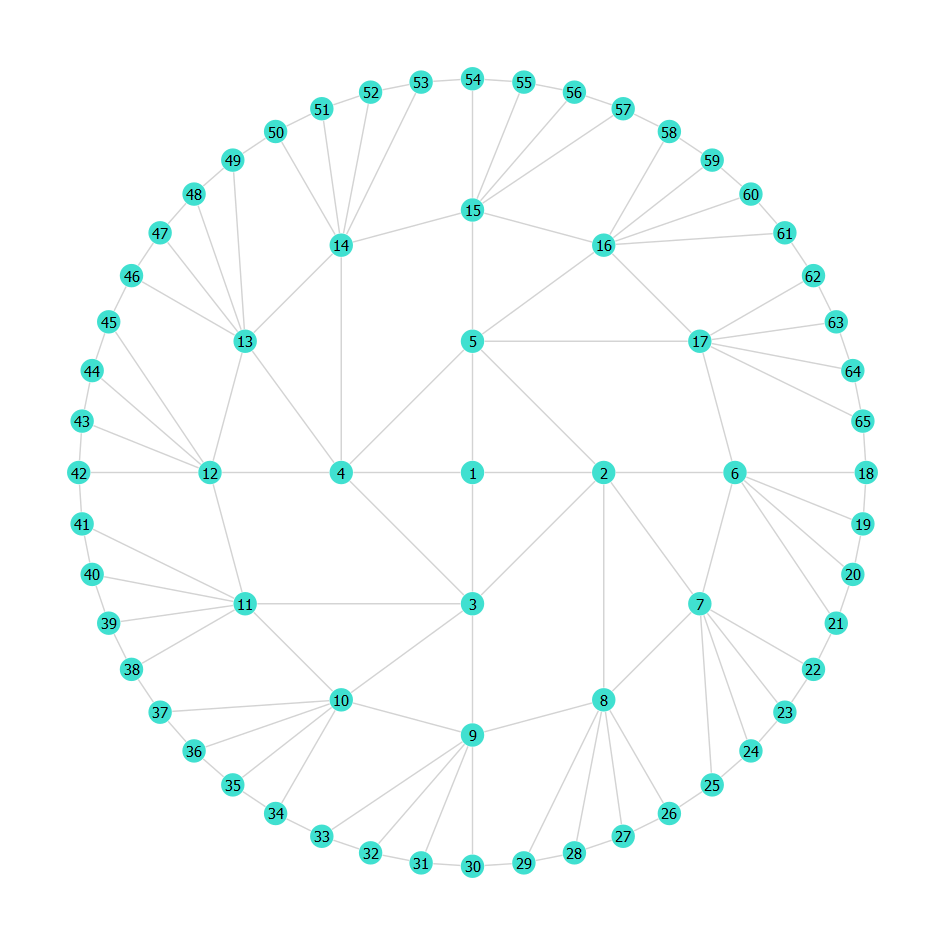}&(h) \includegraphics[width=40mm,height=30mm]{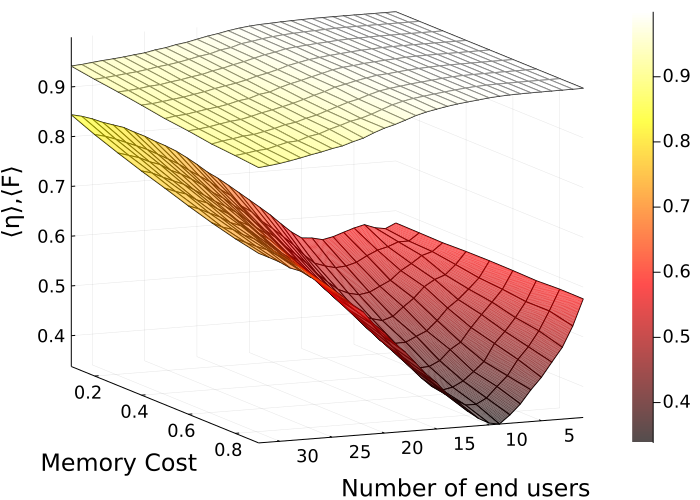}
    \end{tabular}
         (i)\includegraphics[width=40mm,height=30mm]{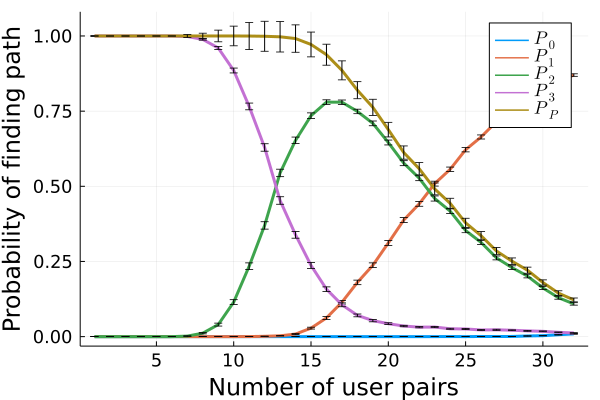}
    \caption{Three different edge schemes in connected tree topology are presented to make three different topologies. Each Topology has $65$ nodes, $128$ edges each with the weight of $0.1 dB$ for both loss and dephasing, and the network statistics were collected for $500$ trials. Each user pair was allowed to use $3$ paths to send their entangled pairs, and they could utilise $3$ different temporal layers. (c), (f) and (i) are plotted for path probabilities and path purification behaviour. In network performance plots (b), (e), and (h), the lower surface represents efficiency, and the upper surface represents fidelity.}
    \label{topo-scal}
\end{figure}

We use a connected tree topology of 65 nodes to show scalability, with $2(65-1)=128$ edges. Fig.~\ref{topo-scal}(a) is a connected tree with just one depth that has 64 nodes connected to one central node Fig.~\ref{topo-scal}(d) has two depths, the first one with $8$ nodes while the second one has a branch parameter of $7$, and Fig.~\ref{topo-scal}(g) has three depths with branching ratios of $4, 3$ and $4$. To present the effects of topologies, all test networks are simulated with each edge having two weights/costs, `loss' and `Z (dephasing)', with an arbitrary value of $0.1 \text{dB}$. Each network is tested with three temporal layers, where costs of loss and dephasing from utilizing quantum memory vary arbitrarily from $0.1 \text{dB}$ to $0.9 \text{dB}$. Each user pair can use three paths to route their entangled pairs.
The probability of purifying entangled pairs from multiple paths
\begin{align}
    P_P=\frac{\sum_{n\geq2}P_n}{(1-P_0)}.
\end{align}
depends on the probability of using $n$ number of paths $P_n$ and the probability of finding zero paths $P_0$.

One depth in Fig.~\ref{topo-scal} (a), (b) and (c) is the idealistic case, showing no competition, and every user pair gets to route using multiple paths. This is impractical due to
many factors like very small area coverage and extreme high usage of a central node. As we increase the depth of the connected
tree it becomes more practical allowing more user pairs to route via multiple paths. It can be observed from Figs.~\ref{topo-scal} (b), (e), and (h), that the average efficiency between the number of user pairs decreases till the purification threshold. This reflects the ability of a graph to accommodate multiple user pairs, with most of them opting for multi-path routing. Pairs must use single-path routing at the user threshold to accommodate maximum user pairs. The increase in efficiency after the purification threshold indicates that user pairs are resorting to shortest-path routing. Any more user pairs will have higher chances of not finding any path to route their entangled pairs.  Figs.~\ref{topo-scal} (b), (e) and (h) show that when we increase the depth, the average efficiency of
the network decreases meaning that more user pairs are routed via multi-path. while Figs.~\ref{topo-scal} (f) and (i) reflect that more depths result in greater user accommodation. 
\begin{table}[!ht]
    \centering
    \begin{tabular}{c|c|c|c}
         Depth & Branches & Purification Threshold & User Threshold \\ \hline
         1 & 64 & none & none\\
         2 & 8,7 & 10 & 27\\
         3 & 4,3,4 & 13 & 29\\
         4 & 3,2,3,2 & 15 & 32
    \end{tabular}
    \caption{This table shows how the different edge schemes affect network thresholds.}
    \label{threshold table}
\end{table}

The observations from Fig.~\ref{topo-scal} are more evident from Table.~\ref{threshold table}, showing purification and user thresholds for different depths. From Table.~\ref{threshold table}, it is clear that networks with more depths give better outcomes in a connected tree topology. The edge scheme of a graph is essential not only for network performance but also for finding paths between user pairs. Fig.~\ref{topo-scal}(c) shows that the edge scheme of Fig.~\ref{topo-scal}(a) is highly beneficial for accommodating all the user pairs. This edge scheme links any two nodes by traversing through only two edges. The additional advantage of temporal routing in three layers makes it suitable for user accommodation. Path-finding in Fig.~\ref{topo-scal}(d) and (g) is different from Fig.~\ref{topo-scal}(a) because they have extra depths. Here the user pairs compete to find paths as shown in Fig.~\ref{topo-scal}(f) and (i). However, the network with three depths has more significant thresholds for purification and user accommodation at the expense of low efficiencies.
\subsection{Impact of Network Topology on Q-Network Statistics}\label{3c}
In the previous section, we discussed the effects of different edge schemes on network statistics. Here a comparison is presented between two fundamental topologies that are inherently different regarding several edges and edge schemes, including temporal routing, to simulate the effect of quantum memories. This comparison has been made with an understanding that a lattice due to its specific arrangement of edges has an inherent advantage in routing problems. To make a fair comparison, a network of 64 users is assumed in two different topologies: a connected tree with $4$ branches and $126$ edges in Fig.~\ref{3,2,3,2 CT stats}(a), and an $8 \cross 8$ lattice (grid) with $112$ edges in Fig.~\ref{3,2,3,2 CT stats}(b). All other variables like edge costs, number of trials, temporal layers, memory costs, and number of allowed paths per user pair are kept constant, as defined in Sec.~\ref{3B}.
\begin{figure}[!ht]
    \centering
    \begin{tabular}{c c}                        (a)\includegraphics[width=36mm,height=36mm]{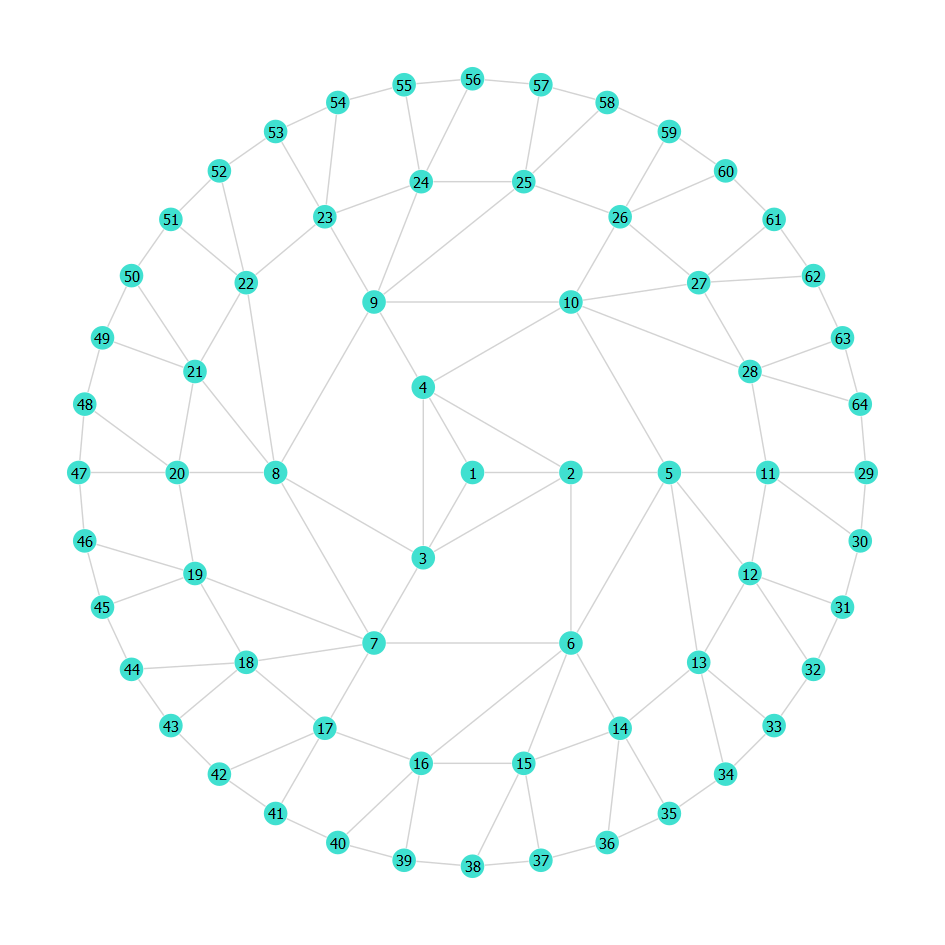}&  (b)\includegraphics[width=34mm,height=34mm]{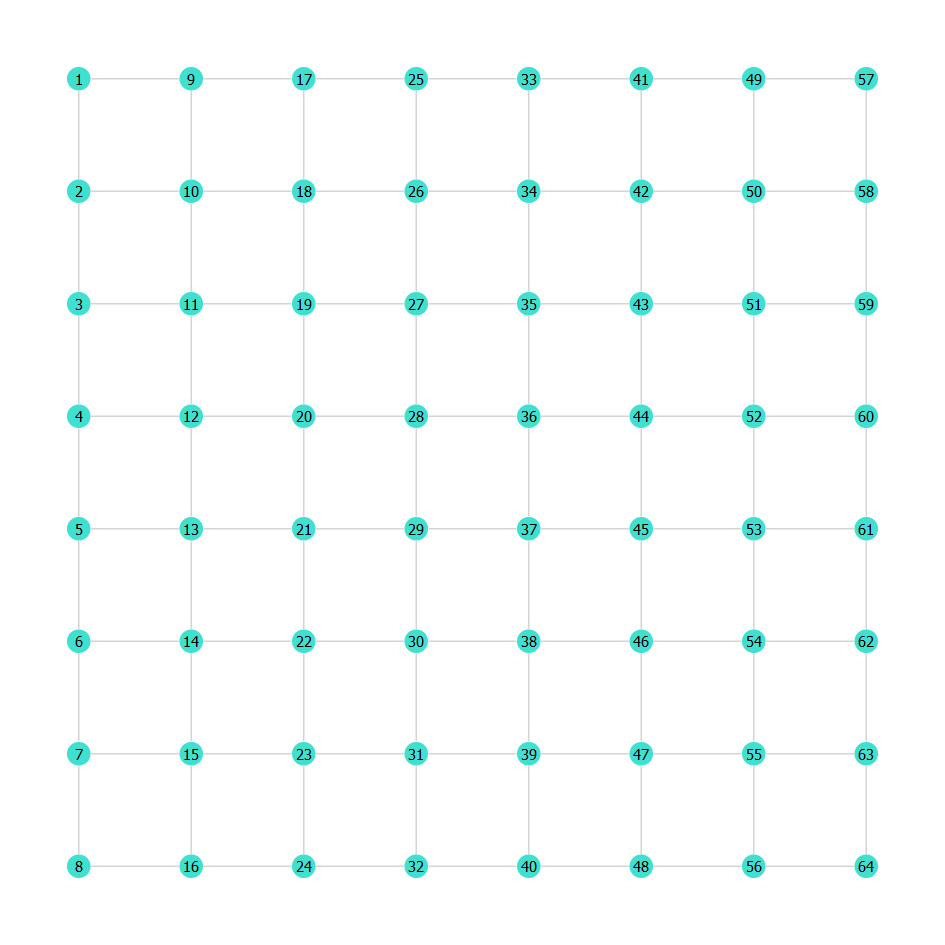}\\(c) \includegraphics[width=35mm,height=30mm]{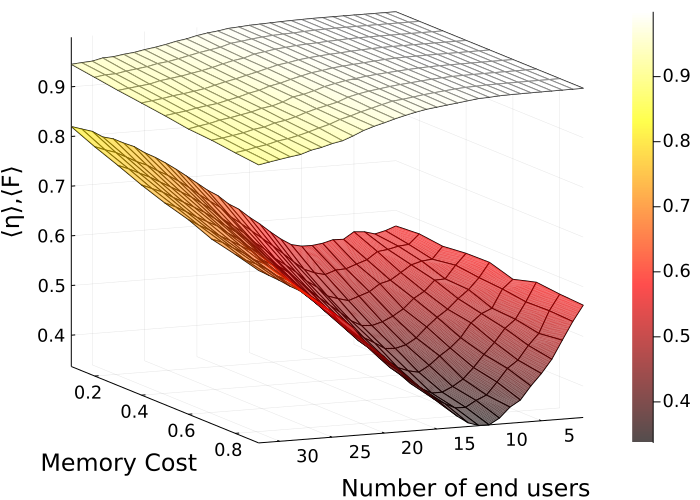}& (d)\includegraphics[width=35mm,height=30mm]{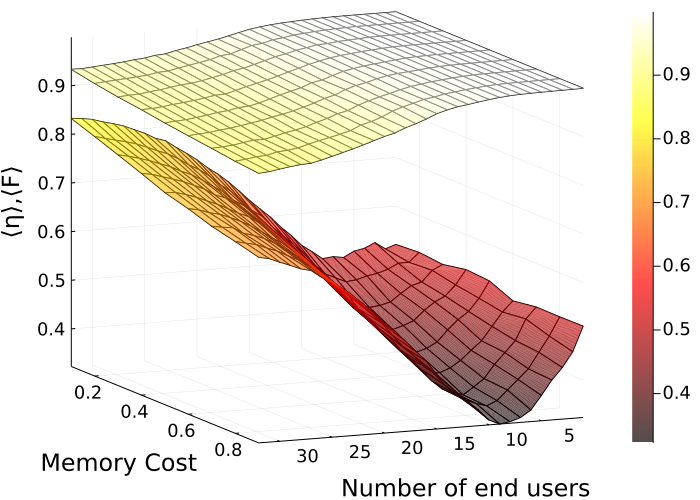}\\(e)\includegraphics[width=35mm,height=30mm]{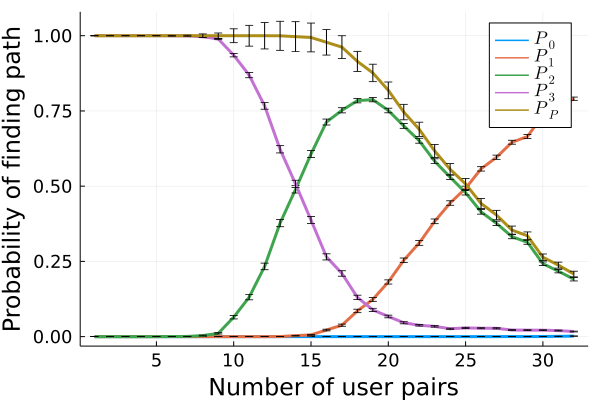}
    & (f)\includegraphics[width=35mm,height=30mm]{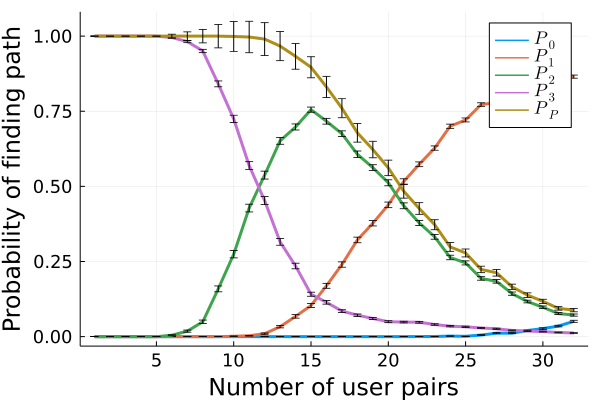}\\
    (g)\includegraphics[width=35mm,height=30mm]{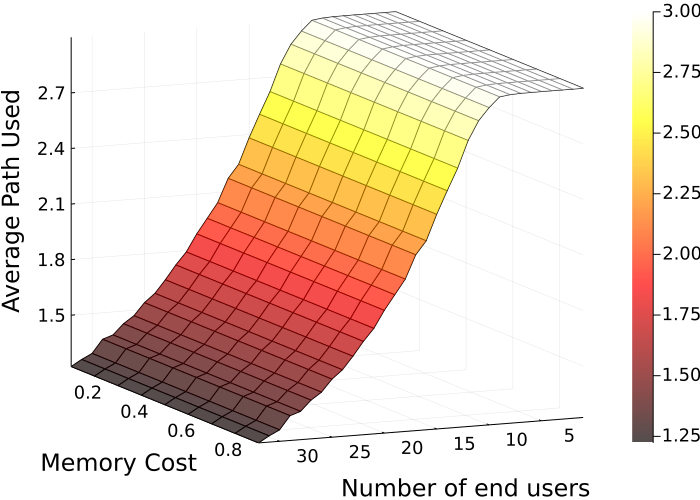}&(h) \includegraphics[width=35mm,height=30mm]{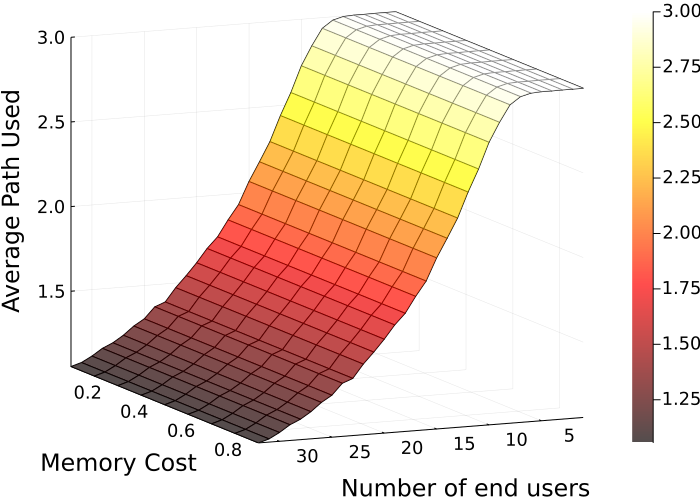}
    \end{tabular}
    \caption{Path routing and network statistics for 64 node topologies. One is a connected tree with 126 edges in four depths with branching parameters of 3,2,3, and 2, respectively, and the other is a lattice network with 112 edges. (a),(b) shows the edge scheme in the network, (c) (d) shows the average performance of the network versus user competition and memory cost, where the lower surface presents efficiency, and the upper surface represents average fidelity between user pairs. (e),(f) shows an account of probabilities for path opting strategy versus user competition, and (g) (h) shows average paths used by each user pair versus user competition.}
    \label{3,2,3,2 CT stats}
\end{figure}
As observed from Figs.~\ref{3,2,3,2 CT stats}(c) and (d), the average network performance of a connected tree is slightly better than that of a lattice. The efficiency surface graph of the connected tree is less steep than the lattice's, which indicates that the network is more resilient toward user competition and provides more options to route entangled pairs to each user pair. Multi-path routing, which provides a good gain in fidelity between end users, also worsens the net efficiency between each user pair~\cite{leone2021qunet}. 

A slight advantage of average performance between the connected tree and lattice, shown in Figs.~\ref{3,2,3,2 CT stats}(c) and (d), is supported by Figs.~\ref{3,2,3,2 CT stats}(e) and (f). The plots show path probabilities for each user pair scenario, which translates towards user accommodation and competition. We can see in Figs.~\ref{3,2,3,2 CT stats}(c) and (e), that for almost 15 user pairs the connected tree provides multiple paths to all users, resulting in high fidelity among them. Afterwards, we see that due to competition majority of user pairs resort to single-path routing. It is worth that for the connected tree, in a high competition scenario, all 32 user pairs are being accommodated as the probability of finding zero path is zero. Whereas, in Figs.~\ref{3,2,3,2 CT stats}(b) and (d), we see that lattice is providing multiple paths to only 12 user pairs. In high competition environment, we see that after the first 25 user pairs, finding even a single path for some user pairs is difficult. Thus, the lattice lags to accommodate as many user pairs as a connected tree.

This depicts that the connected tree is more tolerant and accommodates more user pairs with a higher chance of providing multiple paths than lattice.
The average path supports these observations based on probabilities visualized in Figs.~\ref{3,2,3,2 CT stats}(g) and (h). On average, three paths are used by seven user pairs in the lattice and nine in the connected tree, two paths by 16 user pairs in the lattice, and 19 in the connected tree. Both networks provide at least one path to all 32 user pairs, provided that networks have three temporal layers.

\subsection{Impact of Network Topology on Secret Key Rates in QKD}\label{3D}
 As shown earlier, networks behave differently when scaled to higher user competition environments. The aim is to see how the QKD secret key rates are affected by network topology, comparing results for a lower number of user pairs preferring multi-path routing to a higher number of user pairs that are more prone towards single-path routing.  Now for applicability, networks shown in Sec.~\ref{3c} are made test beds for quantum communication via distributing a secret key between each user pair through quantum key distribution protocol. An ideal approach is used where the users perform operations that do not add decoherence. Thus, all the errors can be accounted for by the paths used.

To calculate the secret key rate, we use the raw qubit rate~\cite{leone2021qunet},
\begin{align}
    R=\frac{M(1-P_0)\eta'}{\tau},
\end{align}
where $M$ is the number of user pairs in the network, and $P_0$ is the probability of the network failing to provide a single path between a user pair. $\eta'$ is the transmission efficiency of the path/s used by each user pair to send their entangled pairs given in Eq.~\ref{6}. Lastly, $\tau$ is the number of temporal layers available to user pairs to ensure maximum success in finding a route. As for the QKD secret key rate per user pair, we use~\cite{Nemoto2016}
\begin{align}
    \nonumber C&=R(1-\mathcal{H}(\varepsilon))\\
    &=\frac{(1-P_0)\eta^{'}}{\tau}\Big[1-\mathcal{H}(F')\Big].
\end{align}
Here $\mathcal{H}(F')=F'\log_2(F')+(1-F')\log_2(1-F')$ is the entropy and $F'$ is the final fidelity of all the paths traversed for each pair. Networks are simulated to observe the effects of user competition and network topology on QKD rates.
\begin{figure}
    \centering
    \begin{tabular}{c c}
        (a)\includegraphics[width=35mm,height=35mm]{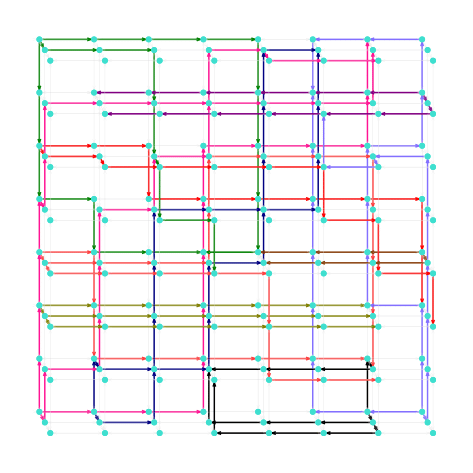}&\includegraphics[width=40mm,height=35mm]{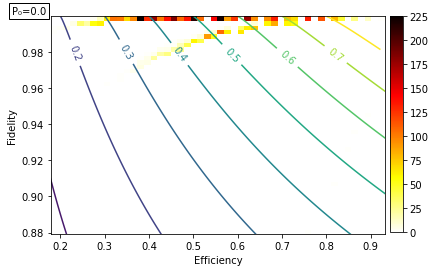}\\(b)\includegraphics[width=35mm,height=35mm]{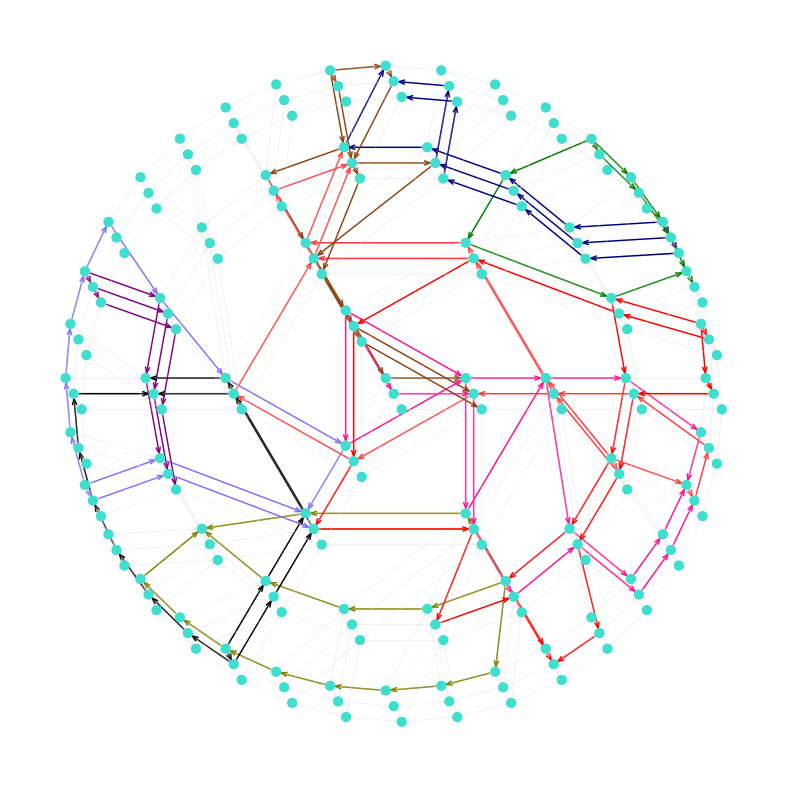}     &\includegraphics[width=40mm,height=35mm]{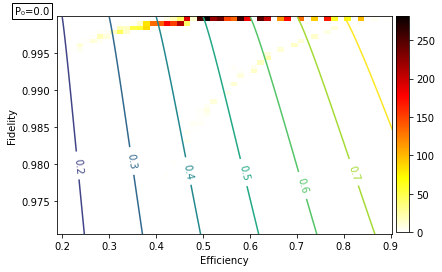}\\(c)\includegraphics[width=35mm,height=35mm]{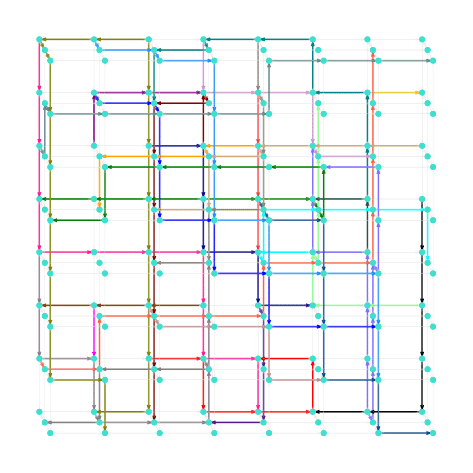} & \includegraphics[width=40mm,height=35mm]{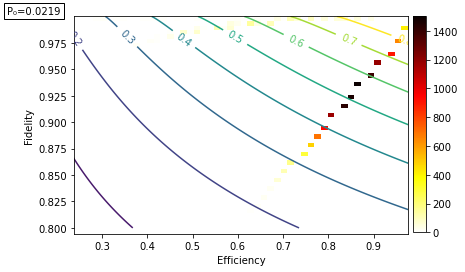}\\(d)
        \includegraphics[width=35mm,height=35mm]{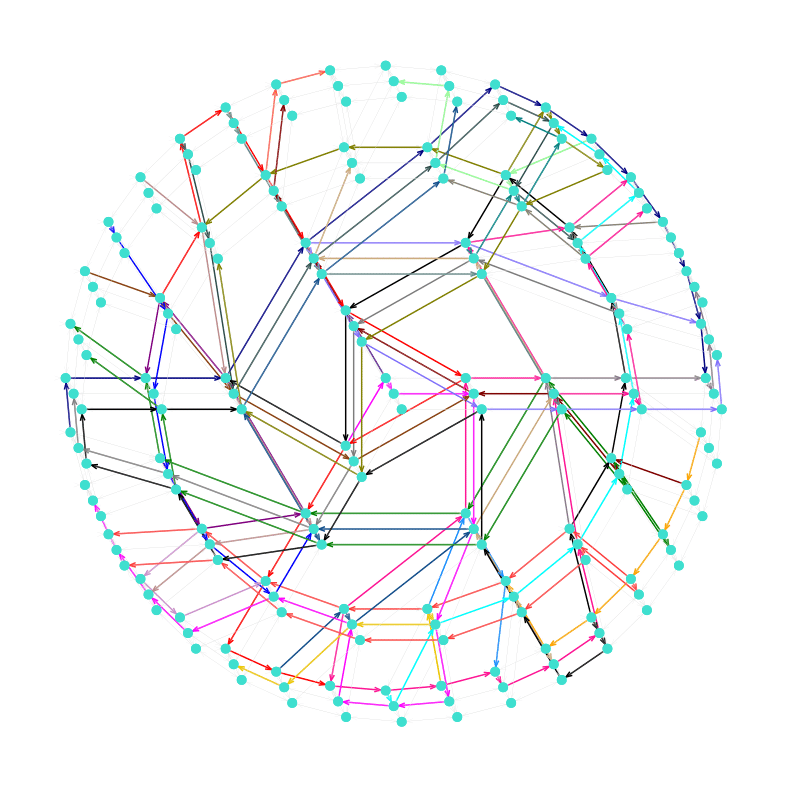}& \includegraphics[width=40mm,height=35mm]{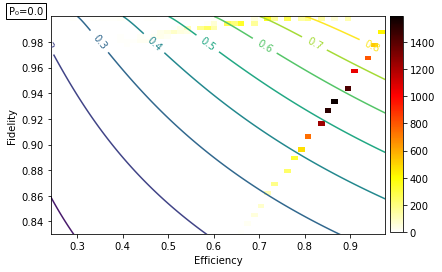}
    \end{tabular}
    \caption{Network topologies with three temporal layers are shown adjacent to their graphical representation of routing data opted by user pairs in 500 trials, with an overlay of contours depicting QKD secret key rates. The networks are simulated for low user competition of 10 user pairs for (a) $8\cross 8$ lattice and (b) $4$ depth connected tree and for high user competition of 30 pairs for (c) $8\cross 8$ lattice and (d) $4$ depth connected tree}
    \label{QKD Lat-CT}
\end{figure}
The user competition and topology of the simulated network affect the path selection between user pairs, affecting the secret key generation rates.

It can be observed from the routing data shown in Fig.~\ref{QKD Lat-CT}(a) that out of 10 user pairs, some choose to use two rather than three paths. Efficiencies of paths in the lattice are spread over a wide range, with few being more significant than others. This means user pairs far apart in lattice have greater costs for multi-path routing. Therefore, such pairs can distribute a secret key through QKD but will have to succumb to lower key rates. Conversely, some user pairs are close and can share a key with a very high key generation rate. Most of the routing data gives key rates below $0.5$ due to the very high-efficiency cost, indicating that the poor choice of paths provided by the network decreased the output key rate. On the other hand, the connected tree in a low competition environment, shown by Fig.~\ref{QKD Lat-CT}(b), benefits from the topology to provide a shorter path between user pairs. Hence most of the routing data have better efficiency costs. The key rates generated by them are better than $0.5$ even with multi-path routing.

In high-user competition, both networks behave somewhat similarly. Still, the lattice has a slight chance of almost $2\%$, where it can fail to provide a path between user pairs, thus affecting the overall purpose of networking. From Fig.~\ref{QKD Lat-CT}(c), it is observed that when lattice becomes saturated with user pairs, it prefers single path routing with optimized routes, thus providing better efficiency costs resulting in improved key rates. For example, the connected tree topology shown by Fig.~\ref{QKD Lat-CT}(d), where it was easier to optimize routing paths, maintains the trend of better paths for QKD rates in single path routing as well, with a few user pairs having the luxury of multi-path routing.
\subsection{Simulating Memory-less Entanglement Distribution Networks for QKD with Redundant Edges}\label{3E}
Multi-path routing and network topology play a crucial role in future multi-user entanglement distribution networks with memories, to accommodate the desired number of user pairs with feasible efficiency-fidelity trade-off. However, efficient quantum memories are not readily available. Here we show a clear advantage of using redundant edges in practical memory-less networks as well. We consider two significantly different networks for $14$ universities in Islamabad, the capital city of Pakistan.
 
In real-world networks, each edge is not the same. Some will be longer than others, affecting the costs of utilizing those edges. So for the Islamabad network, we made the edge costs dependent on the shortest distance between the two nodes that the edge joins, along with a base cost of $0.1 \text{dB}/\text{km}$. One of the two networks simulated for Islamabad is a minimal spanning tree (MST) with $13$ edges. This is one of the most common topologies used for multi-node P2P networks. By definition, it is a graph that joins all the nodes with a minimum number of edges. The other is a network with $91$ edges based on a complete graph that connects each node with every other node, thus providing at least one path to each user pair at any given moment. 
\begin{figure}
    \centering
    \begin{tabular}{c c}
        \includegraphics[width=38mm,height=35mm]{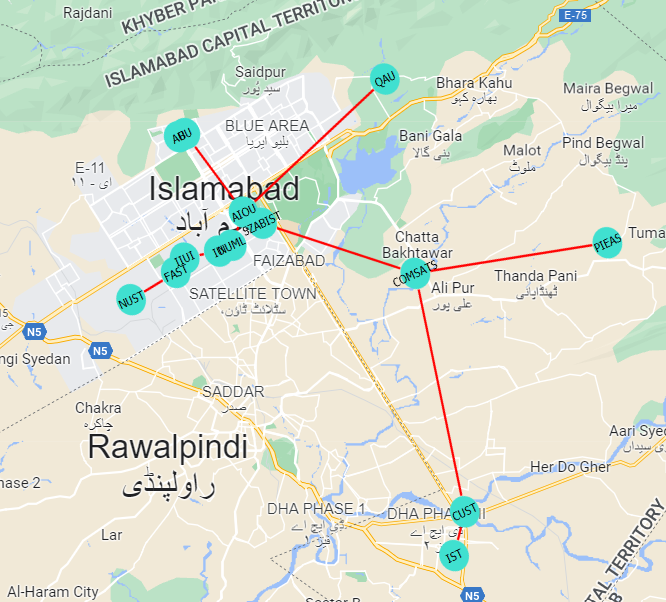} &(a)
         \includegraphics[width=40mm,height=35mm]{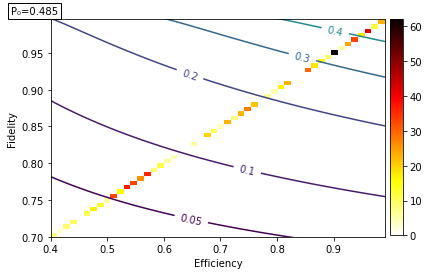}\\
         \includegraphics[width=38mm,height=35mm]{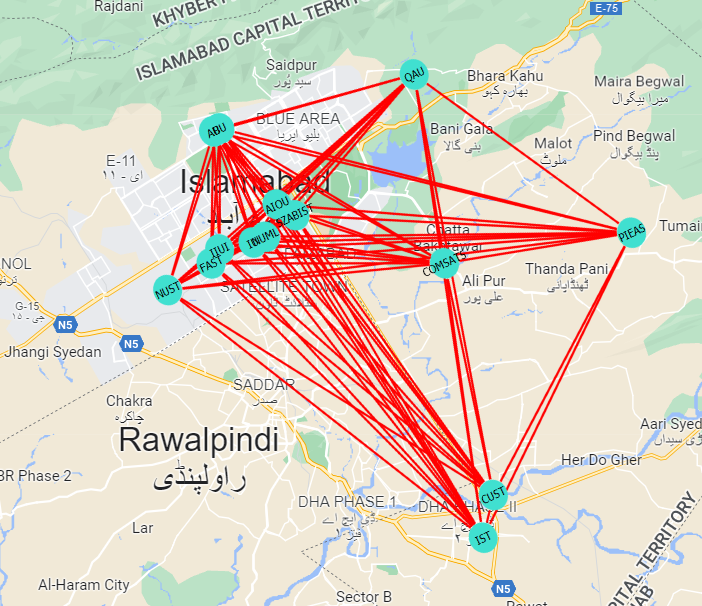} &(b)
         \includegraphics[width=40mm,height=35mm]{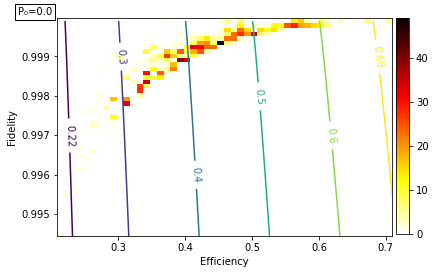}
    \end{tabular}
    \caption{(a) A minimal spanning tree network that has a $61.37 km$ total optic fiber channels, (b) A connected graph network that has a $1109.67 km$ total optic fiber channels. Efficiency-Fidelity routing path data overlayed with QKD secret key rate contours showing a linear distribution for the trade-off. The network was simulated for $500$ trials and $3$ user pairs. It was found to fail almost $49\%$ of the time to provide a single path to the user pair.}
    \label{MST ISB}
\end{figure}

An MST-like topology will likely fail around 49\% times in providing a path to anything greater than one user pair. Therefore, when simulated for $500$ trials, the routing path data between 3 user pairs is linearly distributed. Grouping in data might be due to the higher cost for longer paths between far-away user pairs. That utilises many channels at lower efficiency-fidelity regions, and shorter paths between closer or neighboring user pairs have auspicious values for fidelity and efficiency. The high chance of failure makes MST a terrible candidate for a multi-user entanglement distribution network. This means that even though user pairs might be neighboring, there is a strong chance that those channels are already under use by some other user pair, rendering it useless. Thus, meager secret key rate generation rates were expected from Fig.~\ref{MST ISB}(a).

The complete graph (CG) projects a different trade-off than MST. The CG network has around  $91$ channels with a very high net length which guarantees multi-path routing leading to low collective path efficiencies but high fidelities. The key generation rate per user pair improves because all three user pairs have many options for path routing. This improvement causes the complete graph to surpass MST in QKD key rate at very low efficiencies, as seen in Fig.~\ref{MST ISB}(b).
\section{Conclusion}\label{4}
For robust near-term quantum internet, the study proposes a novel approach to use network topologies with redundant edges to overcome the need for efficient quantum memories. To achieve high user accommodation and improved path-finding we suggest using thin connected trees and for minimum dependence on quantum memories, we present a complete tree as an extreme case of edge redundancy. To conclude, the findings presented in this study highlight the significance of topology, edge scheme, and multi-path routing in the design and optimization of entanglement distribution networks for scalable quantum internet applications. The research focused on comparing the performance of connected tree and lattice networks with different edge schemes in terms of multi-user routing, path optimization algorithms, and quantum key distribution (QKD) rates. The simulations demonstrated that a connected tree topology with redundant edges outperforms the acyclic tree structure by providing multiple paths between a larger number of user pairs. By maximizing the depth of the primary tree network, the connected tree topology offers increased routing options at the cost of some channel losses. This was supported by the analysis of network statistics, which revealed an increasing trend of purification and user thresholds with increased depth in the connected tree network.

The study also emphasizes how crucial network topology is to path optimization techniques. Because of the number of channels crossed, the lattice network showed poor average path efficiency, particularly in situations where user competition was minimal. In terms of average path efficiency and the capacity to create various routes between user pairs, the connected tree topology performed better. Due to its ability to establish entanglement over several pathways, multi-path routing had a substantial impact on QKD rates, resulting in greater secret key rates per user pair. Because the connected tree architecture may give more routing alternatives while utilizing fewer channels, it produced higher QKD rates. The practical ramifications of these findings for quantum networking engineers and scientists are noteworthy since they offer vital insights into the optimization and allocation of resources.

\section{Acknowledgements}
The authors thank and acknowledge Dr Peter P. Rohde for many fruitful discussions and helpful insights.

\bibliography{ref.bib}

\end{document}